\documentclass[submission,copyright,creativecommons]{eptcs}
\providecommand{\event}{PDMC 2011} 
\usepackage{breakurl}             

\usepackage{amsmath,amsfonts,amssymb}
\usepackage{color}
\usepackage{graphicx}
\usepackage{xspace}
\usepackage{comment}
\usepackage{times}
\usepackage{multicol}
\usepackage{program}

\newcommand{\define}[1]{\emph{#1}}
\newcommand{\Kind}{\textrm{\textsc{Kind}}\xspace}
\newcommand{\KindInv}{\textrm{\textsc{Kind}-\textsc{Inv}}\xspace}
\newcommand{\PKind}{\textrm{\textsc{PKind}}\xspace}
\newcommand{\Mpi}{\textrm{\textsc{Mpi}}\xspace}
\newcommand{\bool}{\ensuremath{\mathsf{bool}}\xspace}
\newcommand{\inte}{\ensuremath{\mathsf{int}}\xspace}
\newcommand{\real}{\ensuremath{\mathsf{real}}\xspace}

\renewcommand{\vec}[1]{\mathbf{#1}}
\newcommand{\lo}{\ensuremath{\mathcal{L}}\xspace}

\newcommand{\sys}{\ensuremath{\mathcal{S}}\xspace}

\def\REPEAT{\qtab\keyword{repeat}\ }
\def\UNTIL{\untab\keyword{until}\ }

\def\ENDPROC{\untab}

\def\COMMENT#1{\texttt{// #1}}
\def\keyword#1{\mbox{\normalshape\bf #1}}

\def\ENDWHILE{\untab}

\newcommand{\bassert}{\mathsf{assert}_\mathrm{base}}
\newcommand{\iassert}{\mathsf{assert}_\mathrm{ind}}
\newcommand{\bentail}{\mathsf{entailed}_\mathrm{base}}
\newcommand{\ientail}{\mathsf{entailed}_\mathrm{ind}}
\newcommand{\breset}{\mathsf{reset}_\mathrm{base}}
\newcommand{\ireset}{\mathsf{reset}_\mathrm{ind}}
\newcommand{\bcex}{\mathsf{cex}_\mathrm{base}}
\newcommand{\icex}{\mathsf{cex}_\mathrm{ind}}

\title{\PKind: A parallel k-induction based model checker\thanks{Work partially supported by AFOSR grant \#FA9550-09-1-0517 and NSF grant \#1049674.}}
\author{Temesghen Kahsai
\institute{The University of Iowa}
\email{temesghen-kahsaiazene@uiowa.edu}
\and
Cesare Tinelli
\institute{The University of Iowa}
\email{cesare-tinelli@uiowa.edu}
}

\def\titlerunning{PKind}
\def\authorrunning{T. Kahsai \& C. Tinelli}

\begin{document}
\maketitle

\begin{abstract}
  \PKind is a novel parallel $k$-induction-based model checker 
  of invariant properties
  for finite- or infinite-state Lustre programs. Its architecture, which is
  strictly message-based, is designed to minimize synchronization
  delays and easily accommodate the incorporation of incremental
  invariant generators to enhance basic $k$-induction. 
  We describe \PKind's functionality and main features, and
  present experimental evidence that \PKind significantly speeds up
  the verification of safety properties and, due to incremental
  invariant generation, also considerably increases the number of
  provable ones.
\end{abstract}

\section{Introduction}

\PKind is a parallel model checker based on the
$k$-induction principle, used to verify 
invariant
 properties of
programs written in the specification/programming language
Lustre~\cite{lustre}. Lustre is a synchronous data-flow language
that operates on infinite streams of values of three basic types: \bool,
\inte (finite precision integers), and \real (floating point numbers).
%
\PKind assumes an idealized version of Lustre, which treats \inte as the
type of mathematical integers, and \real as the type of rational
numbers.  
Its reasoning about Lustre programs is done in the context of 
a first-order quantifier-free logic 
that includes uninterpreted functions and mixed real-integer linear arithmetic. 
Idealized Lustre programs can be faithfully and readily encoded 
as transition systems in this logic  (see~\cite{hagen08} for more
details).  
\PKind relies on the SMT solvers CVC3~\cite{BarTin-CAV-07} and
Yices~\cite{DutDeM-RR-06}, in alternative, 
as satisfiability solvers for this logic.

\PKind's architecture is strictly message-based and 
designed to minimize synchronization delays and easily accommodate the
concurrent automatic generation of invariants to bolster basic
$k$-induction. A first level of parallelism is introduced in the
$k$-induction procedure itself by executing the base and the inductive
steps concurrently. A second level 
allows the addition
of one or more independent processes that incrementally generate
\emph{auxiliary} invariants for the system being verified. These invariants are fed to
the $k$-induction loop as soon as they are produced and used to
strengthen the induction hypothesis. 

To the best of our knowledge, this sort of parallel architecture has
not been presented in previous work on parallel model checking. Our
approach is orthogonal to those in previous work~\cite{BBCR10b} that
focus on other sources of parallelism, including parallelization
across the processes of an asynchronous transition system. Most
closely related to ours is the work by E{\'e}n \textit{et
  al.}~\cite{ES2003} who describe a sequential implementation of
SAT-based $k$-induction in which a 
Bounded Model Checking 
loop is interleaved with one
performing just the inductive step of $k$-induction.
Our approach goes beyond that work, not only in using a genuinely
parallel architecture, but also by incorporating concurrent invariant
generation processes. Another line of
related
work is exemplified
by~\cite{Erika07,tarmo09,Bradley11}, which discuss a different type of
parallelism in the BMC algorithm. There, satisfiability checks are
done concurrently within the SAT solver. This 
too is
orthogonal
to our approach, as the parallelism we exploit is not at the level of
the underlying solver, but at the level of the $k$-induction procedure.

In the current version of \PKind, invariant generation is achieved
using a novel incremental version of an offline invariant discovery
scheme we developed in previous work~\cite{Kahsai-Ge-Tinelli-10}. This
general scheme consists in sifting through a large set of formulas
generated automatically from a transition system's description,
looking for possible invariants.  The formulas in the set, the
\define{candidate invariants}, are all instances of a template encoding a
decidable relation over the system's data
types. In~\cite{Kahsai-Ge-Tinelli-10}, a single invariant is generated
at the end of the process as a conjunction of template instances, each
of which is $k$-inductive for some $k$.  In contrast, in the version
developed for \PKind, instances that are $k$-inductive for the same
$k$ are discovered and returned before instances that are
$k'$-inductive for some $k' > k$.

Before describing \PKind's architecture, 
we briefly recall the definition of $k$-induction~\cite{She00}.
Assume a logic $\lo$ and a transition system $S$ specified in the logic
by an initial state condition $I(\vec x)$ and a two-state transition 
relation $T(\vec x,\vec x')$
where $\vec x, \vec x'$ are vectors of state variables.
A state property $P(\vec x)$ is invariant for $S$,
i.e., satisfied by every reachable state of $S$,
if the following entailments hold in $\lo$ for some $k \geq 0$:
%
\begin{gather}
I(\vec{x_0}) \land  T(\vec{x_0},\vec{x_{1}}) \land \cdots \land T(\vec{x_{k-1}},\vec{x_k}) \models P(\vec{x_0}) \land \cdots \land P(\vec{x_k}) 
\label{eq:base} 
 \\ 
 T(\vec{x_0},\vec{x_{1}}) \land \cdots \land T(\vec{x_k},\vec{x_{k+1}}) \land P(\vec{x_0}) \land \cdots \land P(\vec{x_k}) \models P(\vec{x_{k+1}})
 \label{eq:step} 
\end{gather}
%
A counterexample trace for the base case entailment (1) indicates that 
the property 
$P$ 
is falsified in a reachable state of $S$, 
and so is not invariant. 
A counterexample trace for the induction step entailment (2) does not
provide the same information because it may start from an unreachable state.  
The normal way to try to rule out such spurious counterexamples is 
to increase the depth $k$ of the induction.  This is, however, 
not guaranteed to succeed because some invariant properties are not
$k$-inductive for any $k$.  
An additional, and orthogonal, way of enhancing $k$-induction is 
to strengthen the induction hypothesis 
$P(\vec{x_0}) \land \cdots \land P(\vec{x_k})$
with the assertion of previously proven invariants for $S$.\footnote{
Further improvements involve the addition of \emph{path compression}
constraints and checks~\cite{She00,deMRS-CAV-03},
which not only speed up computation 
but also guarantee completeness in certain cases---including all finite state systems.
}

\section{PKind's architecture}
 
\PKind is implemented in OCaml and is built with components from
(the sequential) \Kind model checker~\cite{hagen08} and the \KindInv invariant
generator~\cite{Kahsai-Ge-Tinelli-10} we developed in previous work.
Its concurrent components are implemented as operating system
processes since OCaml's concurrency model 
does not take advantage of multi-processor hardware to parallelize 
thread-level computation.  
We used the \Mpi
(\textit{Message Passing Interface}) API~\cite{mpi97} to implement
communication among the different processes.

\begin{figure}[t] 
\centering
\includegraphics[scale=0.34]{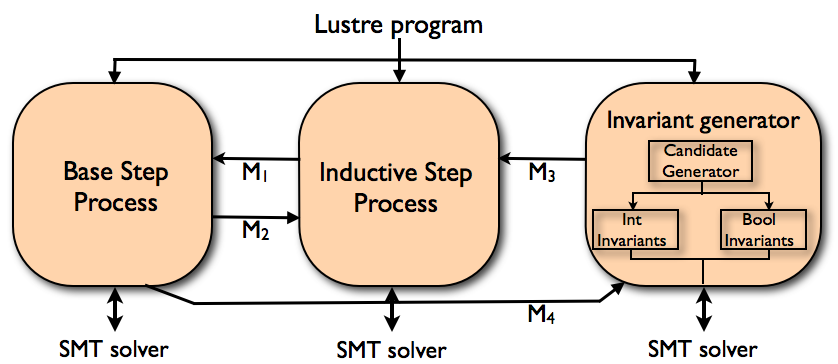} 
\caption{
\PKind general architecture.
}
\label{fig:pkind_arch}
\end{figure}

Figure 1 
illustrates the general architecture of
\PKind.  For simplicity, we consider only the case of one invariant
generation process.  The extension to an arbitrary number of invariant
generators (of the same type) is straightforward.  Each process uses
its own copy of the SMT-solver; and each of them receives as input the
formulas encoding the Lustre program to be checked.  The three
processes, described below, cooperate as in the Actor model,
exchanging messages asynchronously by means of non-blocking
$\mathsf{send}$ and $\mathsf{receive}$ operations on message queues.
\medskip

\noindent
\textbf{Base process:}
Starting from $k=0$, the base process checks the
  entailment of case~(\ref{eq:base}) in the $k$-induction definition,
  for increasing values of $k$.  If the entailment fails, it produces
  a counterexample and sends a termination signal $M_2$ to the
  inductive step process.  Otherwise, it keeps increasing $k$ and
  checking the entailment until it receives a message $M_1$ from the
  inductive step process stating that the process has
  proven the inductive step~(\ref{eq:step}) for a certain value $n$
  of $k$.  At that point, the base case process checks that the base
  case holds for that value of $k$, succeeding if it does and
  returning a counterexample otherwise.
\medskip

\noindent
\textbf{Inductive step process:}
  This process checks the inductive step entailment for increasing
  values of $k$ until one of the following occurs, in this order:
  $(i)$ the entailment succeeds, $(ii)$ it receives a termination
  signal from the base case process, or $(iii)$ it receives an
  invariant from the invariant generation process.  In the
  latter case, the inductive step process asserts the discovered
  invariant for all the states involved so far (from $0$ to $k+1$)
  before repeating the process with the same value of $k$. The inductive
  step process has an auxiliary role in this architecture---it is the base case process
  that determines whether the property is $k$-inductive or not.

\medskip

\noindent
\textbf{Invariant generation process:} 
This process is composed of three modules. The \textit{Candidate
  generator}, synthesizes a set of candidate invariant from predefined
templates.  The \textit{Int} and \textit{Bool Invariants} modules
generate integer and Boolean type invariants respectively.
Invariants
are sent to the inductive step process, in messages $M_3$, as soon as
they are discovered. The process keeps sending newly discovered
invariants until it processes all candidates or it receives a
termination signal $M_4$ from the base process.

This process produces $k$-inductive invariants for a given transition
system \sys from a \define{template} $R[\_,\_]$, 
a formula of
$\lo$ representing a decidable binary relation over one of system's
data types.  The invariants are conjunctions of instances $R[s,t]$ of
the template produced with terms $s,t$ from a set $U$ of terms over
the state variables of \sys.  The set $U$ can be determined
heuristically from \sys (and possibly also from the property $P$ to be
proven) in any number of ways.\footnote{ In our experiments, we
  constructed 
  $U$
  with terms occurring in the $\lo$-encoding of \sys 
  plus
  a few distinguished constants from the domain of \sys's
  variables.  }

\begin{figure}[t]
\small
\begin{minipage}[t]{.45\linewidth}
\begin{program}
\PROC |invariant\_generator| \BODY
  C = \bigwedge_{s,t \in U} R[s,t]
  k = -1
  \breset; \ireset
  \COMMENT{Phase 1}
  \REPEAT
     k = k + 1
     \bassert(T_k)
     \mathit{changed} = \mathsf{false}
     \WHILE \lnot\,\bentail(C_k) \DO
        \mathit{cex} = \mathsf{project}(\bcex,C_k)
        C = \mathsf{filter}(C,\mathit{cex})
        \mathit{changed} = \mathsf{true}
     \ENDWHILE
  \UNTIL \lnot\,\mathit{changed}
  \COMMENT{Phase 2}
  \iassert(T_1 \land C_0 \land \cdots \land T_{k+1} \land C_k)
  \WHILE \lnot\,\ientail(C_{k+1}) \DO 
    \mathit{cex} = \mathsf{project}(\icex,C_{k+1})
    C = \mathsf{filter}(C,\mathit{cex})
  \ENDWHILE
  \mathsf{send}(D, \mathsf{ind\_proc})
\ENDPROC
\end{program}
\begin{center}
\textbf{Version A} \\
(non-incremental)
\end{center}
\end{minipage}
\qquad
\begin{minipage}[t]{.4\linewidth}
\begin{program}
\PROC |invariant\_generator| \BODY
  C = \bigwedge_{s,t \in U} R[s,t]
  k = -1
  \breset
  \REPEAT
     k = k + 1
     \bassert(T_k)
     \WHILE \lnot\,\bentail(C_k) \DO 
        \mathit{cex} = \mathsf{project}(\bcex,C_k)
        C = \mathsf{filter}(C,\mathit{cex})
     \ENDWHILE
  D = C
  \ireset
  \iassert(T_1 \land D_0 \land \cdots \land T_{k+1} \land D_k)
  \mathit{changed} = \mathsf{false}
  \WHILE \lnot\,\ientail(D_{k+1}) \DO 
    \mathit{cex} = \mathsf{project}(\icex,D_{k+1})
    D = \mathsf{filter}(D,\mathit{cex})
    \mathit{changed} := \mathsf{true}
  \ENDWHILE
    \mathsf{send}(D, \mathsf{ind\_proc})
  \UNTIL \lnot\,\mathit{changed}
\ENDPROC
\end{program}
\begin{center}
\textbf{Version B} \\
(incremental)
\end{center}
\end{minipage}

\caption{
\textbf{General scheme for invariant generation.}
The procedures $\mathsf{reset}$ and $\mathsf{assert}$ and 
the functions $\mathsf{entailed}$ and $\mathsf{cex}$, 
are all indexed by the copy of the $\lo$-solver used by the process.
$\mathsf{reset}$ empties the solver's set of asserted formulas;
$\mathsf{assert}(F)$ adds the formula $F$ to that set;
$\mathsf{entailed}(F)$ returns $\mathsf{true}$
iff the currently asserted formulas $\lo$-entail $F$.
When invoked after a call to $\mathsf{entailed}$ that returned
$\mathsf{false}$, $\mathsf{cex}$ returns a counterexample 
for that failed entailment.
The function $\mathsf{project}$ 
takes an assignment $\alpha$ for state variables
and a formula $F$, and
returns the restriction of $\alpha$ to the variables of $F$.
The function $\mathsf{filter}$ takes a conjunctive property $P$
and an assignment $\alpha$ for $P$'s variables, and 
returns the property obtained from $P$
by removing all conjuncts falsified by $\alpha$.
In Version B, $D$ is meant to be a copy of $C$, not a renaming.
The call $\mathsf{send}(D, \mathsf{ind\_proc})$
sends $D$ to the inductive step process.
}
\label{fig:invariant_generation}
\end{figure}

\subsection{Invariant Generation}

%
The general invariant generation scheme introduced in~\cite{Kahsai-Ge-Tinelli-10}
is illustrated in Figure~\ref{fig:invariant_generation}, Version A.
The two-phase procedure, also based on $k$-induction,
starts with a conjunction $C$ of candidate invariants
and first eliminates from $C$ as many conjuncts as possible that have 
an actual counterexample.
Then it attempts to prove the resulting $C$ $k$-inductive,
pruning any unprovable conjunct until it succeeds---possibly with an empty $C$
in the worst case.
All conjuncts of $C$ that remain at the end are invariant and can be returned.
Note that in Phase 1 the search for counterexamples stops when,
for some $k$, $C$ is falsified by no $k$-reachable states.  This is
just a heuristic termination condition which does not preclude the
possibility that some conjuncts of $C$ may be falsified by longer
counterexamples.  As a consequence, every conjunct of $C$ that does
not pass the test in Phase 2 are conservatively assumed not to be
invariant (even if it may be $k'$-inductive for some $k' > k$) and
removed.

The general invariant discovery scheme above produces a
single conjunctive invariant at the very end.
In a concurrent setting, however, it is better for runtime performance 
to have an incremental invariant generation process, 
which identifies and returns invariants as it goes.
Concretely, it is better for the invariant generation process of \PKind
first to identify, and immediately send to the inductive step process,
conjuncts of $C$ that are $0$-inductive, then identify and send
those that are $1$-inductive, and so on.

Pseudo-code for the incremental procedure is provided 
in Figure~\ref{fig:invariant_generation}, Version B.
The procedure starts again with the conjecture
$C = \bigwedge_{s,t\in U} R[s,t]$.
However, it does the following for every $k \geq 0$.
First, it eliminates from $C$ all conjuncts falsified by a $k$-reachable state. 
Then it makes a copy $D$ of $C$ and tries to prove it $k$-inductive
by checking the $k$-induction step on $D$.
Counterexamples in that step are used to weaken $D$ further,
eliminating more and more conjuncts from it 
until no counterexamples exists. 
The final formula $D$ is $k$-inductive and can be already sent
to the inductive step process.
If $D$ did not need to be weakened at all, it means that 
\emph{all} the conjuncts left in $C$ after the base case check 
were $k$-inductive.
Hence, the whole process terminates.
If $D$ was weakened, the eliminated conjuncts 
are either falsified by a longer counterexample or 
possibly $k$-inductive for a larger $k$.
Those conjuncts are still in $C$, so in that case 
the procedure increases $k$ by 1 and repeats with the larger $k$.

We point out that 
the conjuncts left in $D$ (and sent out because $k$-inductive) are
not removed from $C$ before repeating.
This is for simplicity but also for convenience 
because on the new iteration they will end up be in $D$ again,
strengthening the induction hypothesis for the inductive step check.
That means, however, that the set of conjuncts sent out 
at iteration $k+1$ is a (strict) superset of the one sent out at
iteration $k$. 
This can be remedied by keeping the previous $D$ and
sending out only the difference between the new $D$ and the old.\footnote{
In our experimental evaluation, however, this enhancement 
did not seem necessary 
because in most cases the first few sets of invariants were 
enough for the process $\mathsf{ind\_proc}$ to prove 
the input property.
}

\vspace{-0.4cm}
\subsection{Additional Features}

\PKind takes full advantage of all the sophisticated features of its
two embedded SMT solvers, such as their being on-line, incremental,
backtrackable, and able to compute and return satisfying assignements
or unsatisfiable cores of input formulas.  
It provides three
different running modes:
in \texttt{\small k\_induct} mode, \PKind
functions as a basic parallel $k$-induction model checker, scheduling
only the processes for the base and the inductive step;
in
\texttt{\small no\_inc\_invariant} mode, it creates also the invariant
generation process. The invariant generation implemented here is
Version A of Figure~\ref{fig:invariant_generation};
in \texttt{\small
  inc\_invariant} mode, the invariant generation is done in an
incremental fashion, as in Version B of
Figure~\ref{fig:invariant_generation}.

\PKind includes a number of optional enhancements on top of this basic
procedure,
inherited from the \Kind checker.
The main ones are: \textit{path compression},
\textit{termination checking} and \textit{abstraction}.  Path
compression strengthens the induction hypothesis with a formula that,
in essence, removes from consideration program executions with
repeated states or more than one initial state.  Termination checking
allows it to prove a non-$k$-inductive property in some
cases by realizing that the reachable state space has been completely
explored.  Abstraction generates a structural abstraction of the
idealized Lustre program and performs something similar to a CEGAR
loop in both the base case and the induction step, with the goal of
improving the tool's scalability.  
See~\cite{hagen08} for more
information on these and other minor enhancements.

\vspace{-0.4cm}
\section{Experimental evaluation}

We have evaluated \PKind and its different working modes experimentally
against \Kind and \KindInv
using the same benchmark set as in~\cite{Kahsai-Ge-Tinelli-10}.\footnote{ Tools and experimental data can be found at
  \url{http://clc.cs.uiowa.edu/Kind}.} 
That set collects a variety of benchmarks from several sources,
with each benchmark consisting of a Lustre program 
and a property to be checked for invariance.

Let us call a benchmark \emph{valid} if its property holds 
in every reachable state of the associated program, 
and \define{invalid} otherwise. \Kind is able
to show 438 of the 941 benchmarks in our set invalid by
returning a (separately verified) counter-example trace for the
program.  \Kind reports 309 of the remaining benchmarks as valid and
diverges on the remaining 194 benchmarks, even with very large timeout
values. 

For the experiments described here the benchmark set is divided in two
groups: the 503 valid or unsolved benchmarks, and the 438
invalid ones.  Our tests compare \PKind vs.~\Kind
along two dimensions: \emph{precision}, measured as the percentage of solved
benchmarks, and \emph{runtime}.  For the latter, we were interested
specifically in evaluating how the parallel architecture speeds up
$k$-induction in the case of invalid benchmarks, and how
\emph{incremental} invariant generation contributes in decreasing
solving times for valid benchmarks. The experiments were run on a
6-core 2.67 GHz Intel Xeon machine, with 6 GB of physical memory,
under RedHat Enterprise Linux 4.0.  Version 1.0.9 of the Yices solver
was used both for \PKind and \Kind.

\subsection{Precision results}

As in~\cite{Kahsai-Ge-Tinelli-10}, we used \KindInv to generate
invariants offline.  We first ran \KindInv on
the benchmark set for 200 seconds.  For each of the benchmarks where
\KindInv did not time out, we obtained a set of invariants, and added
them to the program as a single conjunctive assertion.  The asserted
formula was the constant \true\ when \KindInv timed out or ended up
discarding all conjectures from the initial set.  Then, we
ran \Kind in inductive mode with a 100s timeout on each
invariant-enhanced benchmark.  
\Kind's precision over the 503 valid or unsolved benchmarks 
grows from 61\%, without invariants, to 85\%, with the added invariants.  
For each additional benchmark proved valid, 
the property goes from (most likely) not $k$-inductive for any $k$ 
to $k$-inductive for $k \leq 16$.

Using the same time out value, in \texttt{\small k\_induct} mode \PKind
proves valid the same benchmarks proven by \Kind with no invariants.
In either \texttt{\small no\_inc\_invariant} and \texttt{\small inc\_invariant}
mode, \PKind proves the same benchmarks proven by \Kind with
invariants, despite the fact that it generates invariants on the fly,
whereas \Kind gets them as input.\footnote{ As a precaution against
  implementation errors, each invariant produced by \KindInv or by
  \PKind was independently and successfully verified by giving it to
  \Kind separately as a property to prove for the program from which
  it was generated.  
  }
Precision is also unchanged when comparing \PKind with \Kind run 
(in a much faster) BMC mode on the set of 438 invalid benchmarks.  
Furthermore, in all these cases, solving times are shortened considerably with
\PKind, as discussed below.

\vspace{-0.5cm}

\subsection{Runtime results}

\begin{figure}[t]
\centering
\includegraphics[scale=0.4]{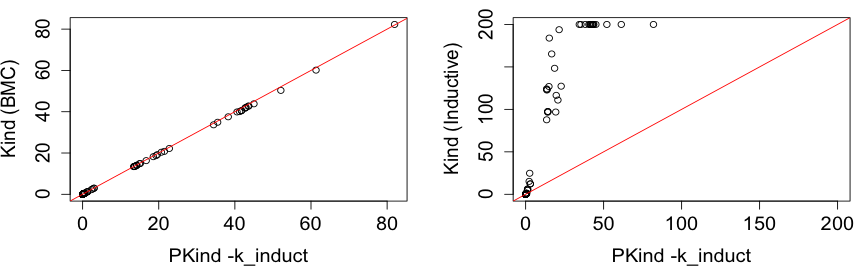} 
\caption{ A comparison of generating counter-examples between \PKind
  \texttt{\small -k\_induct} and \Kind on invalid benchmarks. On the 
  left-hand side we compare \PKind \texttt{\small -k\_induct} and \Kind in
  \textit{BMC} mode. On the right-hand side we
  compare \PKind \texttt{\small -k\_induct} and \Kind in
  \textit{inductive} mode.  }
\label{fig:invalid}
\end{figure}

The two scatter plots in Figure~\ref{fig:invalid} compare
the runtimes of \PKind \texttt{\small -k\_induct} with those of \Kind
respectively in BMC mode (left-hand side plot),
which performs just bounded model checking,
and inductive mode (right-hand side plot),
which performs full $k$-induction.
The timeout for both tools was 200s. 
As the first plot clearly shows,
the runtimes of the two tools are pretty much the same in the first case. 
Due to some overhead of the \Mpi implementation in \PKind, 
for some benchmarks \Kind is slight faster. 
The average solving time for \PKind \texttt{\small -k\_induct} is 2.34s,
while for \Kind in \textit{BMC} mode it is 2.29s.
The superiority of \PKind is clearly illustrated with the second plot.
For a more quantitative comparison, 
both tools are able to produce a counterexample in less than 1s
per benchmark for 90\% of all invalid benchmarks.
However, \PKind \texttt{\small -k\_induct} takes less than 82s to solve
any of the invalid benchmarks,
whereas \Kind in inductive mode 
times out on half of them.
It is worth stressing that \PKind \texttt{\small -k\_induct} 
is also superior from a usability perspective 
because it combines the speed of \Kind's BMC mode over invalid inputs
with the generality and precision of \Kind's inductive mode,
without requiring the user to chose between these two modes.

Moving to \PKind's invariant generating modes,
since invariants for \Kind are generated offline,
it is not possible to do direct runtime comparisons 
between the two systems over the valid or unsolved benchmarks.
%
A more interesting comparison is between 
the two invariant generating modes themselves.
In a nutshell, our results show that 
the addition of incrementally produced invariants, 
makes those benchmarks considerably faster to solve.  
In more detail,
we first compared \PKind \texttt{\small -no\_inc\_invariant} and 
\PKind \texttt{\small -inc\_invariant} on all previously valid benchmarks,
those that can be shown valid without the addition of invariants. 
On those benchmarks, the conjunctive invariant produced 
by the non-incremental version has no effect on the $k$-induction loop,
because the property is invariably proven before the invariant is generated.
In contrast, when invariants are generated incrementally, 
they are in many cases sent to the inductive step process
early enough to play a role in reducing the overall solving time.


\begin{figure}[t]
\centering
\includegraphics[scale=0.4]{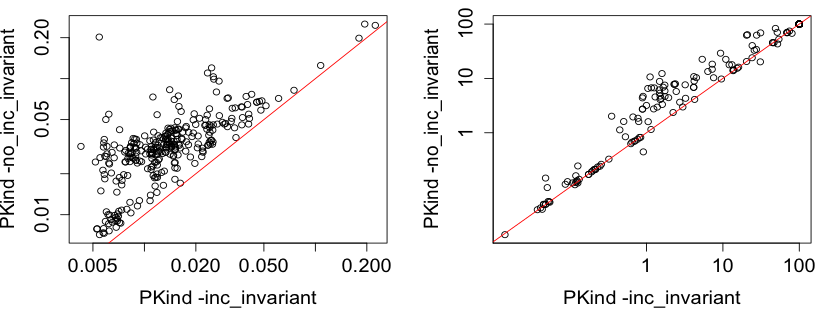} 
\caption{ A comparison of solving the benchmarks between \PKind
  \texttt{\small -no\_inc\_invariant} and \PKind
  \texttt{\small inc\_invariant} on valid benchmarks. On the left-hand side
  we compare the two modes on benchmarks which are valid without the
  addition of invariants.  On the right-hand side we compare the two
  tools on the benchmarks previously unsolved.
  Both plots use a log-log scale over seconds.}
\label{fig:incremental}
\end{figure}

The left-hand side of Figure~\ref{fig:incremental} provides a
comparison between the two invariant generation modes. The plot demonstrates how
incremental invariant generation speeds up runtimes. 
Specifically, the incremental version is able to solve
76\% of all the valid benchmarks in less than 0.02 seconds, while the
non-incremental version can do that for only 17\% of those benchmarks.
Looking at the previously unsolved benchmarks, with a timeout of 100s, 
both invariant generation modes solve 66\% of them.  Again, however, the
incremental version is faster than the non-incremental one
although the speed up is less pronounced,
as can be seen in the right-hand side plot of Figure~\ref{fig:incremental}. 
In more detail, the incremental version is able to solve 79\% of all newly
solved benchmarks in less than 9s, whereas the non-incremental version
is able to solve $73\%$ of them.  
It is interesting to observe that in a few
cases non-incremental invariant generation is slight faster than the
incremental one.  The reason for this is that the useful invariant is
one that is $k$-inductive for a relatively larger $k$ than
usual.\footnote{ In most benchmarks in our set, adding $0$- or
  $1$-inductive invariants is enough to make the property provable.  }
In those cases, incrementally sending and processing (useless)
$k$-inductive invariants for lower values of $k$ produces a bit of
overhead.

With regard to the size of our benchmarks, 
the invalid ones range from 200 bytes to 40KB 
of Lustre source code with no comments and standard use of white space.  
The average time to solve an invalid
benchmark of size more than 20KB is 9.12 seconds.  For the valid
benchmarks the size ranges from 400 bytes to 58KB. The average time
to solve a valid benchmark of size more than 20KB bytes is 17.6
seconds. 
Although benchmark sizes may appear small,
we stress that
the majority of these benchmarks are infinite-state systems written in
a high level specification language, so a benchmark's size is not a
good predictor of its difficulty. In fact, looking at runtime data
over the whole benchmark set we found no correlation between benchmark
size and solving times.

\vspace{-0.5cm}

\section{Conclusion and future work}

We have described \PKind, a novel parallel
$k$-induction-based model checker. \PKind is designed to keep
interprocess synchronization overheads to a minimum and easily
accommodate the concurrent automatic generation of invariants 
to bolster the basic $k$-induction procedure.  
While any independent invariant generation techniques 
could be used in principle in our architecture, we have
developed a new one, based on our previous work, that produces
invariants in an incremental fashion to exploit the advantages of the
parallel setting.  
Our experimental results provide an initial and
clear evidence that
the architecture significantly speeds up the verification of safety
properties. In addition, due to the incremental invariant generation, this architecture
considerably increases the number of provable benchmarks.

In future work, we plan to study new ways of automatically generating
invariants for $k$-induction,
both in parallel, i.e., using abstract interpretation techniques,
and on-demand, in response to counterexamples found by the inductive step.
We are also working on a new version of \PKind built from scratch
around the architecture discussed here and 
designed to accept other input languages besides Lustre.

\nocite{*}
\bibliographystyle{eptcs}
\bibliography{pkind}

\begin{thebibliography}{10}
\providecommand{\bibitemdeclare}[2]{}
\providecommand{\urlprefix}{Available at }
\providecommand{\url}[1]{\texttt{#1}}
\providecommand{\href}[2]{\texttt{#2}}
\providecommand{\urlalt}[2]{\href{#1}{#2}}
\providecommand{\doi}[1]{doi:\urlalt{http://dx.doi.org/#1}{#1}}
\providecommand{\bibinfo}[2]{#2}

\bibitemdeclare{article}{Erika07}
\bibitem{Erika07}
\bibinfo{author}{Erika \'{A}brah\'{a}m}, \bibinfo{author}{Tobias Schubert},
  \bibinfo{author}{Bernd Becker}, \bibinfo{author}{Martin Fr\"{a}nzle} \&
  \bibinfo{author}{Christian Herde} (\bibinfo{year}{2011}):
  \emph{\bibinfo{title}{Parallel {SAT} solving in bounded model checking}}
  \bibinfo{volume}{21}(\bibinfo{number}{1}), pp. \bibinfo{pages}{5--21},
  \doi{10.1093/logcom/exp002}.

\bibitemdeclare{inproceedings}{BBCR10b}
\bibitem{BBCR10b}
\bibinfo{author}{J.~Barnat}, \bibinfo{author}{L.~Brim},
  \bibinfo{author}{M.~\v{C}e\v{s}ka} \& \bibinfo{author}{P.~Ro\v{c}kai}
  (\bibinfo{year}{2010}): \emph{\bibinfo{title}{{DiVinE: Parallel Distributed
  Model Checker (Tool paper)}}}.
\newblock In: {\sl \bibinfo{booktitle}{HiBi/PDMC 2010}},
  \bibinfo{publisher}{IEEE}, \doi{10.1109/PDMC-HiBi.2010.9}.

\bibitemdeclare{inproceedings}{BarTin-CAV-07}
\bibitem{BarTin-CAV-07}
\bibinfo{author}{Clark Barrett} \& \bibinfo{author}{Cesare Tinelli}
  (\bibinfo{year}{2007}): \emph{\bibinfo{title}{{CVC3}}}.
\newblock In \bibinfo{editor}{W.~Damm} \& \bibinfo{editor}{H.~Hermanns},
  editors: {\sl \bibinfo{booktitle}{CAV'07}}, \bibinfo{series}{LNCS},
  \bibinfo{publisher}{Springer}, \doi{10.1007/978-3-540-73368-3\_34}.

\bibitemdeclare{inproceedings}{Bradley11}
\bibitem{Bradley11}
\bibinfo{author}{Aaron~R. Bradley} (\bibinfo{year}{2011}):
  \emph{\bibinfo{title}{{SAT}-based model checking without unrolling}}.
\newblock \bibinfo{series}{VMCAI '11}, \bibinfo{publisher}{Springer-Verlag},
  pp. \bibinfo{pages}{70--87}.
\newblock \urlprefix\url{http://dl.acm.org/citation.cfm?id=1946284.1946291}.

\bibitemdeclare{techreport}{DutDeM-RR-06}
\bibitem{DutDeM-RR-06}
\bibinfo{author}{Bruno Dutertre} \& \bibinfo{author}{Leonardo de~Moura}
  (\bibinfo{year}{2006}): \emph{\bibinfo{title}{The {YICES SMT} Solver}}.
\newblock \bibinfo{type}{Technical Report}, \bibinfo{institution}{SRI
  International}.
\newblock \urlprefix\url{http://yices.csl.sri.com/}.

\bibitemdeclare{article}{ES2003}
\bibitem{ES2003}
\bibinfo{author}{Niklas E{\'e}n} \& \bibinfo{author}{Niklas S{\"o}rensson}
  (\bibinfo{year}{2003}): \emph{\bibinfo{title}{Temporal induction by
  incremental {SAT} solving}}.
\newblock {\sl \bibinfo{journal}{Electr. Notes Theor. Comput. Sci.}}
  \bibinfo{volume}{89}(\bibinfo{number}{4}), \doi{10.1016/S1571-0661(05)82542-3}.

\bibitemdeclare{inproceedings}{hagen08}
\bibitem{hagen08}
\bibinfo{author}{George Hagen} \& \bibinfo{author}{Cesare Tinelli}
  (\bibinfo{year}{2008}): \emph{\bibinfo{title}{Scaling up the formal
  verification of {L}ustre programs with {SMT}-based techniques}}.
\newblock In: {\sl \bibinfo{booktitle}{FMCAD '08}},
  \bibinfo{address}{Piscataway, NJ, USA}, pp. \bibinfo{pages}{1--9},
  \doi{10.1109/FMCAD.2008.ECP.19}.

\bibitemdeclare{article}{lustre}
\bibitem{lustre}
\bibinfo{author}{N.~Halbwachs}, \bibinfo{author}{P.~Caspi},
  \bibinfo{author}{P.~Raymond} \& \bibinfo{author}{D.~Pilaud}
  (\bibinfo{year}{1991}): \emph{\bibinfo{title}{The synchronous data-flow
  programming language {L}ustre}}.
\newblock {\sl \bibinfo{journal}{Proceedings of the IEEE}}
  \bibinfo{volume}{79}(\bibinfo{number}{9}), pp. \bibinfo{pages}{1305--1320},
  \doi{10.1109/5.97300}.

\bibitemdeclare{inproceedings}{Kahsai-Ge-Tinelli-10}
\bibitem{Kahsai-Ge-Tinelli-10}
\bibinfo{author}{Temesghen Kahsai}, \bibinfo{author}{Yeting Ge} \&
  \bibinfo{author}{Cesare Tinelli} (\bibinfo{year}{2011}):
  \emph{\bibinfo{title}{Instantiation-Based Invariant Discovery}}.
\newblock In: {\sl \bibinfo{booktitle}{NFM 2011}}, {\sl \bibinfo{series}{LNCS}}
  \bibinfo{volume}{6617}, \bibinfo{publisher}{Springer}, pp.
  \bibinfo{pages}{192--207}.
\newblock \urlprefix\url{http://dl.acm.org/citation.cfm?id=1986308.1986326}.

\bibitemdeclare{inproceedings}{deMRS-CAV-03}
\bibitem{deMRS-CAV-03}
\bibinfo{author}{Leonardo de~Moura}, \bibinfo{author}{Harald Rue{\ss}} \&
  \bibinfo{author}{Maria Sorea} (\bibinfo{year}{2003}):
  \emph{\bibinfo{title}{Bounded Model Checking and Induction: From Refutation
  to Verification}}.
\newblock In: {\sl \bibinfo{booktitle}{CAV 2003}}, {\sl \bibinfo{series}{LNCS}}
  \bibinfo{volume}{2725}, \bibinfo{publisher}{Springer},
  \doi{10.1007/978-3-540-70952-7\_21}.

\bibitemdeclare{book}{mpi97}
\bibitem{mpi97}
\bibinfo{author}{Peter~S. Pacheco} (\bibinfo{year}{1997}):
  \emph{\bibinfo{title}{Parallel programming with MPI}}.
\newblock \bibinfo{publisher}{Morgan Kaufmann Publishers Inc.}

\bibitemdeclare{inproceedings}{She00}
\bibitem{She00}
\bibinfo{author}{Mary Sheeran}, \bibinfo{author}{Satnam Singh} \&
  \bibinfo{author}{Gunnar St{\aa}lmarck} (\bibinfo{year}{2000}):
  \emph{\bibinfo{title}{Checking Safety Properties Using Induction and a
  {SAT}-Solver}}.
\newblock In: {\sl \bibinfo{booktitle}{FMCAD '00}},
  \bibinfo{publisher}{Springer-Verlag}, \bibinfo{address}{London, UK}, pp.
  \bibinfo{pages}{108--125}, \doi{10.1007/3-540-40922-X\_8}.

\bibitemdeclare{inproceedings}{tarmo09}
\bibitem{tarmo09}
\bibinfo{author}{Siert Wieringa}, \bibinfo{author}{Matti Niemenmaa} \&
  \bibinfo{author}{Keijo Heljanko} (\bibinfo{year}{2009}):
  \emph{\bibinfo{title}{Tarmo: A Framework for Parallelized Bounded Model
  Checking}}.
\newblock In: {\sl \bibinfo{booktitle}{PDMC '09}}, pp. \bibinfo{pages}{62--76},
  \doi{10.4204/EPTCS.14.5}.

\end{thebibliography}

\end{document}